\documentstyle[prc,preprint,tighten,aps]{revtex}
\begin {document}
\draft
\title{
Three-body resonances by complex scaling}
\author{Attila Cs\'ot\'o\cite{email}}
\address{W.~K. Kellogg Radiation Laboratory, 106-38, California
Institute of Technology,\\Pasadena, California 91125, USA}
\date{\today}

\maketitle

\begin{abstract}
\noindent
It is demonstrated that the complex scaling method can be used
in practical calculations to localize three-body resonances.
Our model example emphasizes the fact that in three-body
systems several essentially different asymptotic behaviors
can appear. We show that the possibility of these different
asymptotic configurations can lead to an apparent, resonance like
structure in the three-body continuum.
\end{abstract}
\pacs{PACS numbers: 21.45.+v, 24.30.Gd}

\narrowtext

Three-body resonances, systems which decay into three-body final
states, play important role in few-body physics. The description
of these states are somewhat easier than that of the scattering
processes which lead three (or more) particles in the outgoing
channel. The study of three-body resonances hopefully will help
us to clarify some points of the still not totally solved
scattering problem.

One of the main difficulties in the general formulation of the
quantum mechanical N-body resonance (and scattering) problem
is the prescription of the asymptotic behavior. There is
fundamental difference between the problems where only short
range interactions occur and those where a long range force,
e.g. the Coulomb force, is present \cite{Newton}. There are
methods which eliminate the explicit reference to the unknown
or partially known asymptotics, e.g.\ the J-matrix method
\cite{Yamani}, and the
potential separable expansion method \cite{Gyarmati}, but
up till now they have been developed only for two-body
systems. In the case of bound states the question of the
asymptotic behavior is not a serious problem. The
majority of the methods
work because in most cases the bound state asymptotics
hardly affects the determination of physically
observable quantities. However, in the case of resonances
and scattering states, the asymptotic behavior plays a crucial role.

The complex scaling method (CSM) \cite{Ho} reduces the
description of resonant states to the description of
bound states, thus avoids the problem of asymptotics. This
method handles the non-Coulomb and the Coulomb cases on equal
footing. There is an extension of this method to two-body
scattering states, too \cite{Resigno}. It is an intriguing
question whether
this method can be extended to the general many-body
scattering states. Up till now, three-body resonances have
been investigated e.g.\ using the Faddeev method \cite{Matsui},
by real stabilization \cite{rstab}, and in the time delay
matrix formalism \cite{delay}. In this paper we study them
using the complex scaling method. Although the CSM was used to
describe three-body systems above the three-body breakup
\cite{cs}, the authors did not specify how they
identified three-body resonances.
This research was prompted by the contradicting theoretical
results concerning the existence of a soft dipole resonance in the
neutron halo nucleus $^6$He \cite{Suzuki,Danilin}. What
makes this problem difficult, is the mixture of the underlying
nuclear physics and three-body dynamics. Here we want to clarify
some points of the three-body dynamics.

For the sake of simplicity, we recall the main points of the CSM
in two-body context. In the coordinate space, resonance
eigenfunctions, corresponding to the complex energy solutions of the
\begin{equation}
\widehat H\vert\Psi\rangle=(\widehat T+\widehat
V)\vert\Psi\rangle =E\vert\Psi\rangle
\label{Sch}
\end{equation}
Schr\"odinger equation, show oscillatory behavior in the asymptotic
region, with exponentially growing amplitude, $\sim \exp [i(\kappa
-i\gamma )r]$ $(\kappa ,\gamma >0)$. Thus, they are not elements of
the $L^2$ space. The complex scaling method means that instead of
Eq.\ (\ref{Sch}), we solve the eigenvalue problem of the
transformed Hamiltonian $\widehat
H_\theta=\widehat U(\theta)\widehat H\widehat U^{-1}(\theta)$:
\begin{equation}
\widehat H_\theta\vert\Psi_\theta\rangle=E_\theta\vert
\Psi_\theta\rangle .
\label{Scht}
\end{equation}
$\widehat U(\theta)$ is an unbounded similarity
transformation \cite{Lowdin}, which, in
the coordinate space, acts on a function $f(r)$ such that
\begin{equation}
\widehat U(\theta)f(r)=e^{3i\theta/2}f(re^{i\theta}).
\label{CS}
\end{equation}
(If $\theta$ is real, $\widehat U(\theta)$ means a rotation into the
complex coordinate plane, if it is complex, it means a rotation
and scaling.) The two problems are connected by the
Aguilar-Balslev-Combes theorem \cite{ABC}:
If $\widehat V$ is a (dilation) analytic operator, then
(i) the bound eigenstates of $\widehat H$ are the
eigenstates of $\widehat H_\theta$, regardless of the actual
value of $\theta$, within $0\leq\theta<\pi/2$;
(ii) the continuous spectrum of $\widehat H$ will be
rotated by an angle 2$\theta$;
(iii) a complex generalized eigenvalue of
Eq.\ (\ref{Scht}), $E_{\rm res}=\varepsilon
-i{{1}\over{2}}\Gamma$, $\varepsilon ,\Gamma >0$
(with the wave number $k_{\rm res}=\kappa -i\gamma$,
$\kappa,\gamma >0$),
belongs to the proper spectrum of $\widehat H_\theta$ provided
$2\theta >\vert \arg E_{\rm res}\vert$.
Roughly speaking, the complex scaling transformation changes the
asymptotic wave function from $\exp[i(\kappa -i\gamma)r]$ to
$\exp[i(\kappa -i\gamma)r\exp(i\theta)]$, which, in the case
of $2\theta >\vert \arg E_{\rm res}\vert =2\vert
\arg k_{\rm res}\vert$, makes the diverging wave function
localized.

If we have $N$ particles, we can transform the
problem from one-particle coordinates to certain interparticle
relative coordinates (Jacobi coordinates):
\begin{equation}
\{{\bf r}_1,{\bf r}_2,\dots,{\bf r}_N\}\rightarrow
\{{\bf t}_1,{\bf t}_2,\dots,{\bf t}_{N-1}\},
\label{trans}
\end{equation}
where the origin is fixed at the center of mass, so that
$\sum^N_{i=1}m_i{\bf r}_i=0$, where $m_i$ are the particle
masses. The application of the CSM in this case means that we
transform all relative coordinates under the action of
$\widehat U(\theta)$. Note that the (\ref{trans}) transformation
is linear, therefore if we carry out the complex scaling
transformation in a certain set of Jacobi coordinates, it
results in
the same transformation in all other possible sets of
relative coordinates. It means that we cannot choose the
rotation angles independently, like in the multichannel CSM
\cite{Csotocsm}, in the different configurations.

Our present model problem consists of three particles with
masses $m_1, m_2$, and $m_3$. We choose $m_1=m_2$
so as to have only two different systems of Jacobi coordinates.
The interactions
between the particles are separable forces
\begin{equation}
\widehat V_{ij}=\vert \varphi_0(b)\rangle\lambda_{ij}
\langle \varphi_0(b)\vert , \ \ \ j>i=1,2,3
\label{pot}
\end{equation}
where $\vert\varphi_0(b)\rangle$ is the eigenfunction of
the three dimensional harmonic oscillator with $n=l=0$,
$b$ is the oscillator size parameter, and $\lambda_{ij}$
are the potential strengths. In coordinate space the
interactions depend on the various relative coordinates.
Each interaction has a natural Jacobi coordinate system
in which it has the simplest form, e.g.\ in the (12)3
coordinate system (there is a
relative coordinate between particles 1 and 2, ${\bf t}_{12}$
and another between (1,2) and 3, ${\bf t}_{(12)3}$) the
$V_{12}$ interaction depends only on ${\bf t}_{12}$.

The kinetic energy operator is easy to express in any
coordinate system, e.g.\ in (12)3 it looks like
\begin{equation}
T=-{{\hbar^2}\over{2}}\left [ {{1}\over{\mu_{12}}}
\mbox{\boldmath $\Delta$}_{{\bf t}_{12}}+
{{1}\over{\mu_{(12)3}}}
\mbox{\boldmath $\Delta$}_{{\bf t}_{(12)3}} \right ],
\label{kin}
\end{equation}
where the Laplace operators are differential operators
in the appropriate Jacobians. It is easy to show that
the application of the (\ref{CS}) complex scaling
transformation to our potential and kinetic energy
operators is equivalent with the change of $b$ to
$b\exp (i\theta)$ in (\ref{pot}), and the multiplication
of the right-hand side of (\ref{kin}) by $\exp (-2i\theta)$,
respectively.

As the CSM localizes the resonant wave function, we can
use any bound-state method to describe them. Here we use
the wave function expansion method. We consider three
different wave functions. In coordinate space
\begin{eqnarray}
\Psi_1&=&\sum_{ij}c_{ij}\varphi_i({\bf t}_{12})
\varphi_j({\bf t}_{(12)3}),
\label{psi1}\\
\Psi_2&=&\sum_{ij}d_{ij}\varphi_i({\bf t}_{23})
\varphi_j({\bf t}_{(23)1}),
\label{psi2}
\end{eqnarray}
and the unification of the above two, $\Psi_3=\Psi_1+\Psi_2$.
The expansion coefficients $c$ and $d$ are to be determined
from a variational principle. Our potentials act only
between $s$-waves, therefore each oscillator function
carries zero angular
momentum in (\ref{psi1}) and (\ref{psi2}). We choose the
oscillator size parameter of the wave functions, $\bar b$,
different from $b$ in order
to make the trial functions more flexible. The summation limits
in the wave functions are chosen to reach stable convergence.

For the necessary matrix elements we need to calculate the
overlap of the product oscillator states between different
Jacobi coordinate systems, and the Laplace operator
between such states.
As examples, we show typical terms that appear in the overlap,
kinetic energy, and potential matrix elements:
\begin{equation}
\langle\varphi_i({\bf t}_{23})\varphi_j({\bf t}_{(23)1})
\vert\varphi_k({\bf t}_{12})\varphi_l({\bf t}_{(12)3})
\rangle ,
\end{equation}
\begin{equation}
\langle\varphi_i({\bf t}_{23})\varphi_j({\bf t}_{(23)1})
\vert \mbox{\boldmath $\Delta$}_{{\bf t}_{12}}
\vert \varphi_k({\bf t}_{12})\varphi_l({\bf t}_{(12)3})
\rangle ,
\end{equation}
and
\begin{equation}
\langle\varphi_i({\bf t}_{23})\varphi_j({\bf t}_{(23)1})
\vert\varphi_0({\bf t}_{12})\rangle
\langle\varphi_0({\bf t}_{12})
\vert\varphi_k({\bf t}_{12})\varphi_l({\bf t}_{(12)3})
\rangle.
\end{equation}
\noindent
Using the Talmi-Moshinsky-Tobocman transformation
\cite{Tobocman} we can express a product of oscillator
states, given in a certain Jacobi coordinate system ($\alpha$),
in terms of product
oscillator states in another system ($\alpha '$), e.g.\
\begin{equation}
\varphi^\alpha_i({\bf t}_{23})\varphi^\alpha_j({\bf t}_{(23)1})=
\sum_{k,l} a^{\alpha\alpha'}_{ijkl}
\varphi^{\alpha'}_k({\bf t}_{12})
\varphi^{\alpha'}_l({\bf t}_{(12)3}),
\end{equation}
where the sum is finite, and the transformation coefficients
can be calculated, e.g.\ by the \cite{Bao} program. Using these
transformations, and in addition the overlap between two
oscillator functions with
different size parameters, and the matrix element of the
Laplace operators between such oscillator functions (for the
formulae see e.g.\ \cite{KruppaKato}),
all necessary matrix elements can be calculated analytically.

We choose $m_1=m_2=2$, and $m_3=4$ ($\hbar =1$ and atomic mass
units are used), $b=1.0$, and $\bar b=2.0$. The use of
separable interactions allows us
to set up their strengths in such a way that resonances
occur at prescribed energies in the two-body subsystems
\cite{Csoto}. The choice
$\lambda_{12}=0.6377+i0.0697$ results in a resonance
in the $(1,2)$ subsystem at $E=1.5-i0.5$. The
$\lambda_{13}=\lambda_{23}=1.0$ strengths give a resonance
in the $(1,3)$ and $(2,3)$ subsystems at $1.7553-i0.2438$ energy.
As an illustrative example we show in Fig.\ \ref{fig1}(a)
the result
of a CSM calculation for the $(1,2)$ subsystem. The working
mechanism of the method is clearly seen, the discretized
continuum points are rotated, and the resonance is revealed.

In Figs.\ \ref{fig1}(b)--(d) we show the results of the
three-body CSM calculations using the $\Psi_1$, $\Psi_2$,
and $\Psi_3$ trial functions, respectively. We can see that
as the rotation angles are large enough to localize the resonances
in the subsystems, there are discretized continuum points lying on
strait half-lines which start from the position of the resonances
of the subsystems. These starting points act as non real thresholds.
For example, in Fig.\ \ref{fig1}(b)
the half-line starts at $1.5-i0.5$ which is the resonance energy
in the (12) subsystem. This is in full agreement with the
mathematical theorems \cite{Balslev,Simon}.
In addition to these continuum points, we can
see that there is an isolated point at $4.128-i0.337$ in each figure.
We can identify this point as a three-body resonance. The fact
that this point occur in each figure shows that this state of the
three-body system can show up both $3(12)$ and $1(23)$ asymptotic
behaviors. This is exactly what we expect from a three-body resonance.
This behavior gives us a method to identify
three-body resonances
in practical calculations. But what can we say about the
continuum states which lie on the half-lines starting from the
resonance energies of the subsystems? In these states e.g.\ in the
$1(23)$ configuration there is a resonant state in the $(2,3)$
system and a scattering state between 1 and $(2,3)$. These
continuum states are essentially different from a pure three-body
scattering state which is
represented by the continuum points lying on the half-lines start
at the origin. The resonance+scattering type continuum states can
represent some kind of sequential decay, where the life time of
a quasi-stationary subsystem, $(2,3)$, is longer than the time
needed for 1 and $(2,3)$ to be scattered off.

The fact that different type of continuum states can be present
in a three-body system means that the three-body continuum has
structure in addition to the three-body resonances. Let us
speculate a bit about this additional structure.
In Fig.\ \ref{fig2} we show the distribution of the
continuum points (in 0.2 wide energy bins) of the
model whose wave function is $\Psi_3$. We note here that
there is no qualitative
difference between not complex scaled ($\theta=0$) and complex
scaled results because the CSM causes only a contraction of the
continuum points, while the resonances remain stable. We can see in
Fig.\ \ref{fig2} the resonant structure around 4.2. But, in addition,
we see structures, the concentration of the continuum points,
around 1.6 and 2.0 which values coincide the real parts
of the resonance energies of the subsystems. These resonant-like
structures are apparent and they are the consequence of the fact
that if the energy is larger than the threshold energy of a
subsystem's resonance, a new, resonance+scattering, asymptotic
behavior can appear. We should note that the shape of the
background distribution in Fig.\ \ref{fig2} is surprising.
In the case of two particles, the wave function expansion method
can be considered as if we closed our system into a box whose
size is the finite spatial region of the trial function. This
leads to a $E_n\sim n^2$ $(n=1,2,\dots)$ spectrum. From this, it
follows that the number of continuum points which are in a
$\Delta E$ interval around $E$ (if $\Delta E$ is small) is
$\Delta N\sim 1/\sqrt E \Delta E$. We checked in a two-body model
that this is a really good approximation. Our three-body spectrum
in Fig.\ 2 is, however, strongly differs from such a shape.
In the case of three particles, the Schr\"odinger equation can be cast,
in the hyperspherical coordinate, into a form which is similar to
a two-body equation, see e.g.\ Ref.\ \cite{Fedorov}. If all angular
momenta are zero, in this reformulated Schr\"odinger equation a
centrifugal barrier occures with $L=3/2$. This nonzero $L$ really
leads to a reduction of the low energy solutions, but this is
a marginal decrease. There must be another effect which
supresses the low energy spectrum.

In conclusion, we showed that the complex scaling method can
be used in practical three-body calculations. In this method,
three-body resonances can be identified by those resonant
energy solutions which appear in all Jacobi coordinate system.
We pointed out that the possibility of resonant+scattering type
asymptotic behaviors can lead to an apparent structure in the
three-body continuum. The case of the $^6$He soft dipole
mode is very similar
to our present example. That nucleus is a genuine three-body,
$\alpha +n+n$, system \cite{DanilinPR,Csotohe6}. There are
two resonances
in the $\alpha +n$ subsystem at $0.89-i0.60$ MeV and $4-i4$ MeV
energies \cite{Ajzenberg}. (And there is an antibound state in the
$n+n$ subsystem, which is out of the scope of this article.)
{}From what we can learn our present model, it is fairly possible
that the structure in the $^6$He continuum, which is interpreted
as a fingerprint of a three-body resonance, the so-called soft
dipole resonance, is nothing but a consequence of the three-body
dynamics. Of course here we considered only one side of the
problem. The question of the $^6$He soft dipole resonance can
only be answered in a model which contains both the nuclear
physics and the three-body dynamics properly. The application
of the complex scaling method in a realistic model of $^6$He
is in progress.

\mbox{}

This work was supported by a Fulbright Fellowship (USA), a
Research Fellowship from the Science Policy Office (Belgium)
and by OTKA grant No.\ 3010 (Hungary). I wish to thank Professor
D. Baye, Professor B. Gyarmati and Dr. Z. Papp for useful
discussions.


\widetext
\begin{figure}
\caption{Energy eigenvalues of a complex scaled (a) two-body
problem with $m_1=m_2=2$, and $\lambda_{12}=0.6377+i0.0697$;
(b)--(d) three-body problem with $m_1=m_2=2$, $m_3=4$,
$\lambda_{12}=0.6377+i0.0697$, and
$\lambda_{13}=\lambda_{23}=1.0$.
The trial wave function is: (b) $\Psi_1$, (c) $\Psi_2$, and
(d) $\Psi_3$. The $\theta$ rotation angle is 0.4 rad in each
figure.}
\label{fig1}
\end{figure}

\narrowtext
\begin{figure}
\caption{Distribution of the continuum energy solutions of the
three-body problem specified in Fig.\ \protect\ref{fig1}, with
the trial function $\Psi_3$. The solutions are groupped within
0.2 wide energy bins.}
\label{fig2}
\end{figure}


\begin{references}
\bibitem[*]{email} E-mail address: H988CSO@HUELLA.BITNET \\
On leave from: Institute of Nuclear Research of the Hungarian
Academy of Sciences, P.O.Box 51, Debrecen, H--4001, Hungary
\bibitem{Newton} R.~G. Newton, {\em Scattering theory of waves
and particles}, (Springer-Verlag, New York, 1982).
\bibitem{Yamani} H.~A. Yamani and L. Fishman, J. Math. Phys.
{\bf 16}, 410 (1975).
\bibitem{Gyarmati} B. Gyarmati and A.~T. Kruppa, Phys. Rev.
C {\bf 34}, 95 (1986); Z. Papp, J. Phys A {\bf 20}, 153 (1987);
Phys. Rev. C {\bf 38}, 2457 (1988); and private communication.
\bibitem{Ho} Y.~K. Ho, Phys. Rep. {\bf 99}, 1 (1983);
N. Moiseyev, P.~R. Certain, and F. Weinhold,
Mol. Phys. {\bf 36}, 1613 (1978); Proceedings of the Sanibel
Workshop Complex Scaling, 1978 [Int. J. Quantum Chem. {\bf 14},
343 (1978)]; B.~R. Junker, Adv. At. Mol. Phys. {\bf 18}, 207
(1982); W.~P. Reinhardt, Annu. Rev. Phys. Chem. {\bf
33}, 223 (1982); {\em Resonances--The Unifying Route Towards
the Formulation of Dynamical Processes, Foundations and
Applications in Nuclear, Atomic and Molecular Physics},
edited by E. Br\"andas and N. Elander, Lecture Notes in Physics
Vol. 325 (Springer-Verlag, Berlin, 1989).
\bibitem{Resigno} T.~N. Resigno and P. Reinhardt, Phys. Rev.
A {\bf 8}, 2828 (1973).
\bibitem{Matsui} Y. Matsui Phys. Rev C {\bf 22}, 2591 (1980);
A. Eskandarian and I.~R. Afnan, Phys. Rev. C {\bf 46},
2344 (1992).
\bibitem{rstab} P. Froelich, K. Szalewicz, and R. Stab,
Phys. Lett. A {\bf 129}, 321 (1988).
\bibitem{delay} J.~P. Svenne, T.~A. Osborn, G. Pisent, and
D. Eyre, Phys. Rev. C {\bf 40}, 1136 (1989).
\bibitem{cs} C-Y. Hu and A.~K. Bhatia, Phys Rev. A {\bf 42},
5769 (1990); P. Froelich, A. Flores-Riveros, and S.~A.
Alexander, Phys. Rev. A {\bf 46}, 2330 (1992).
\bibitem{Suzuki} Y. Suzuki, Nucl. Phys. {\bf A528}, 395 (1991).
\bibitem{Danilin} B.~V. Danilin, M.~V. Zhukov, J.~S. Vaagen,
and J.~M. Bang, Phys. Lett. B {\bf 302}, 129 (1993);
L.~S. Ferreira, E. Maglione, J.~M. Bang, I.~J. Thompson,
B.~V. Danilin, M.~V. Zhukov, and J.~S. Vaagen, Phys. Lett.
B {\bf 316}, 23 (1993).
\bibitem{Lowdin} P.~O. L\"owdin, Adv. Quantum Chem. {\bf 19}, 87
(1988).
\bibitem{ABC} J. Aguilar and J.~M. Combes, Commun. Math. Phys.
{\bf 22}, 269 (1971); E. Balslev and J.~M. Combes, {\em ibid.}
{\bf 22}, 280 (1971); B. Simon, {\em ibid.} {\bf 27}, 1 (1972).
\bibitem{Csotocsm} A. Cs\'ot\'o, Phys. Rev. A {\bf 48}, 3390
(1993).
\bibitem{Tobocman} W. Tobocman, Nucl. Phys. {\bf A357},
293 (1981).
\bibitem{Bao} Y. Gan, M. Gong, C. Wu, and C. Bao, Comp. Phys.
Commun. {\bf 34}, 387 (1985).
\bibitem{KruppaKato} A.~T. Kruppa and K. Kat\= o, Prog. Theor.
Phys. {\bf 84}, 1145 (1990).
\bibitem{Csoto} A. Cs\'ot\'o, B. Gyarmati, A.~T. Kruppa, K.~F.
P\'al, and N. Moiseyev, Phys. Rev. A {\bf 41}, 3469 (1990).
\bibitem{Balslev} E. Balslev, in {\em Resonances--Models
and Phenomena}, edited by S. Albeverio, L.~S. Ferreira, and
L. Streit, Lecture Notes in Physics Vol. 211
(Springer-Verlag, Berlin, 1984), p.\ 27.
\bibitem{Simon} B. Simon, Int. J. Quantum Chem. {\bf 14}, 529
(1978).
\bibitem{Fedorov} D.~V. Fedorov, A.~S. Jensen, and K. Riisager,
Phys. Lett B {\bf 312}, 1 (1993).
\bibitem{DanilinPR} M.~V. Zhukov, B.~V. Danilin, D.~V. Fedorov,
J.~M. Bang, I.~J. Thompson, and J.~S. Vaagen, Phys. Rep. {\bf
231}, 151 (1993).
\bibitem{Csotohe6} A. Cs\'ot\'o, Phys. Rev. C {\bf 48}, 165 (1993).
\bibitem{Ajzenberg} F. Ajzenberg-Selove, Nucl. Phys. {\bf A490},
1 (1988).
\end{references}
\end{document}